\documentclass[reprint,amsmath,amssymb,aps,superscriptaddress]{revtex4}
\usepackage[latin9]{inputenc}
\usepackage{graphicx}
\usepackage{esint}
\usepackage{siunitx}

\makeatletter

\DeclareTextSymbolDefault{\textquotedbl}{T1}
\newcommand{\lyxdot}{.}

\@ifundefined{textcolor}{}
{%
 \definecolor{BLACK}{gray}{0}
 \definecolor{WHITE}{gray}{1}
 \definecolor{RED}{rgb}{1,0,0}
 \definecolor{GREEN}{rgb}{0,1,0}
 \definecolor{BLUE}{rgb}{0,0,1}
 \definecolor{CYAN}{cmyk}{1,0,0,0}
 \definecolor{MAGENTA}{cmyk}{0,1,0,0}
 \definecolor{YELLOW}{cmyk}{0,0,1,0}
}

\usepackage[final]{changes}

\usepackage{xr}

\makeatother

\begin{document}

\title{Hyperbolic Phonon Polariton Electroluminescence as an Electronic Cooling Pathway}



\author{E. Baudin}
\email{emmanuel.baudin@lpa.ens.fr}

\affiliation{Laboratoire de Physique de l'Ecole normale sup\'erieure, ENS, Universit\'e
PSL, CNRS, Sorbonne Universit\'e, Universit\'e de Paris, 24 rue Lhomond, 75005 Paris, France}

\author{C. Voisin}

\affiliation{Laboratoire de Physique de l'Ecole normale sup\'erieure, ENS, Universit\'e
PSL, CNRS, Sorbonne Universit\'e, Universit\'e de Paris, 24 rue Lhomond, 75005 Paris, France}

\author{B. Pla\c{c}ais}


\affiliation{Laboratoire de Physique de l'Ecole normale sup\'erieure, ENS, Universit\'e
PSL, CNRS, Sorbonne Universit\'e, Universit\'e de Paris, 24 rue Lhomond, 75005 Paris, France}

\begin{abstract}

Engineering of cooling mechanism is of primary importance for the development of nanoelectronics. Whereas radiation cooling is rather inefficient in nowadays electronic devices, the strong anisotropy of 2D materials allows for enhanced efficiency because their hyperbolic electromagnetic dispersion near phonon resonances allows them to sustain much larger ($\sim 10^5$) number of radiating channels. In this review, we address radiation cooling in 2D materials, specifically graphene hexagonal boron nitride (hBN) heterostructures.
We present the hyperbolic dispersion of electromagnetic waves due to anisotropy, and describe how the spontaneous fluctuations of current in a 2D electronic channel can radiate thermal energy in its hyperbolic surrounding medium.
We show that both the regime of (i) thermal current fluctuations and (ii) out-of-equilibrium current fluctuations can be described within the framework of transmission line theory leading to (i) superPlanckian thermal emission and (ii) electroluminescent cooling.
We discuss a recent experimental investigation on graphene-on-hBN transistors using electronic noise thermometry. In a high mobility semimetal like graphene at large bias, a steady-state out-of-equilibrium situation is caused by the constant Zener tunneling of electrons opening a route for electroluminescence of hyperbolic electromagnetic modes. Experiments reveal that, compared to superPlanckian thermal emission, \replaced{electroluminescent}{electroluminescence} cooling is particularly prominent once the Zener tunneling regime is reached: observed cooling powers are nine orders of magnitude larger than in
conventional LEDs.

\end{abstract}

\maketitle

\section{Introduction}

Heat transfer is a key issue in nano-electronics. This is true for
the silicon MOSFET technology where cooling limits device performance above modest power density $\sim100\si{W.cm^{-2}}$ [\onlinecite{Pop2010nres}] corresponding to a thermal conductance $G_{th}<1\;\si{kW m^{-2}K^{-1}}$.
With the rise of 2D (bi-dimensional) electronics [\onlinecite{Ong20192DM}], this problem becomes
even more  acute  because of the minute heat capacity of the channel
and the reduced thermal links to the environment. In this respect,
the case of graphene-based devices is emblematic : the absence of
a bandgap requires driving graphene transistors in the velocity saturation
regime where Joule dissipation is prominent. In high-mobility devices,
the very limited electron-phonon coupling (nonpolar material, low
defect density) associated to a drop of the Wiedemann-Franz heat conduction
mechanism (due to a vanishing differential conduction in the saturation
regime), yields considerable electronic temperatures, up to several
thousands of kelvins [\onlinecite{Yang2018nnano,Andersen2019science,Ong20192DM}].
In this peculiar situation where phonons remain relatively cold and
thermal conduction is blocked, heat release through radiative processes
becomes a major issue.

It is usually recognized that black-body radiation cooling is rather
inefficient in electronic devices. This essentially stems from the
large impedance mismatch at optical (infrared) frequencies between
vacuum and the electron gas. From a microscopic point of view, this
mismatch is related to the small density of electromagnetic modes
within the light cone as compared to that of the electron gas at high
temperature.
This issue has been \replaced{overcame}{overcome} in recent studies [\onlinecite{Yang2018nnano,Tielrooij2018nnano}],
by using a so-called hyperbolic substrate. Within the Reststrahlen (RS) bands (which roughly corresponds to the optical phonon modes) hyperbolic materials sustain long-distance propagating hyperbolic
phonon-polariton (HPhP) modes that can carry a large momentum [\onlinecite{Biehs2012prl,Biehs2014apl,Caldwell2014ncomm,Dai2015nnano,Kumar2015nl,Giles2016nl,Low2017nmat}]\footnote{\added{for a recent review, see Reference [\onlinecite{Caldwell2019nrevmat}]}}. When the 2D channel becomes coupled to the HPhP modes (through their evanescent field), the whole density of states can be
used to radiate power : this is the so-called super-Planckian regime [\onlinecite{Biehs2014apl,Principi2017prl}].
The enhancement of the radiated power as compared to the regular blackbody can reach up to 5 decades and is not limited by hot phonon effects thanks to the relatively long propagation length of HPhP modes.

\begin{figure}[ht]
\centerline{\includegraphics[width=10cm]{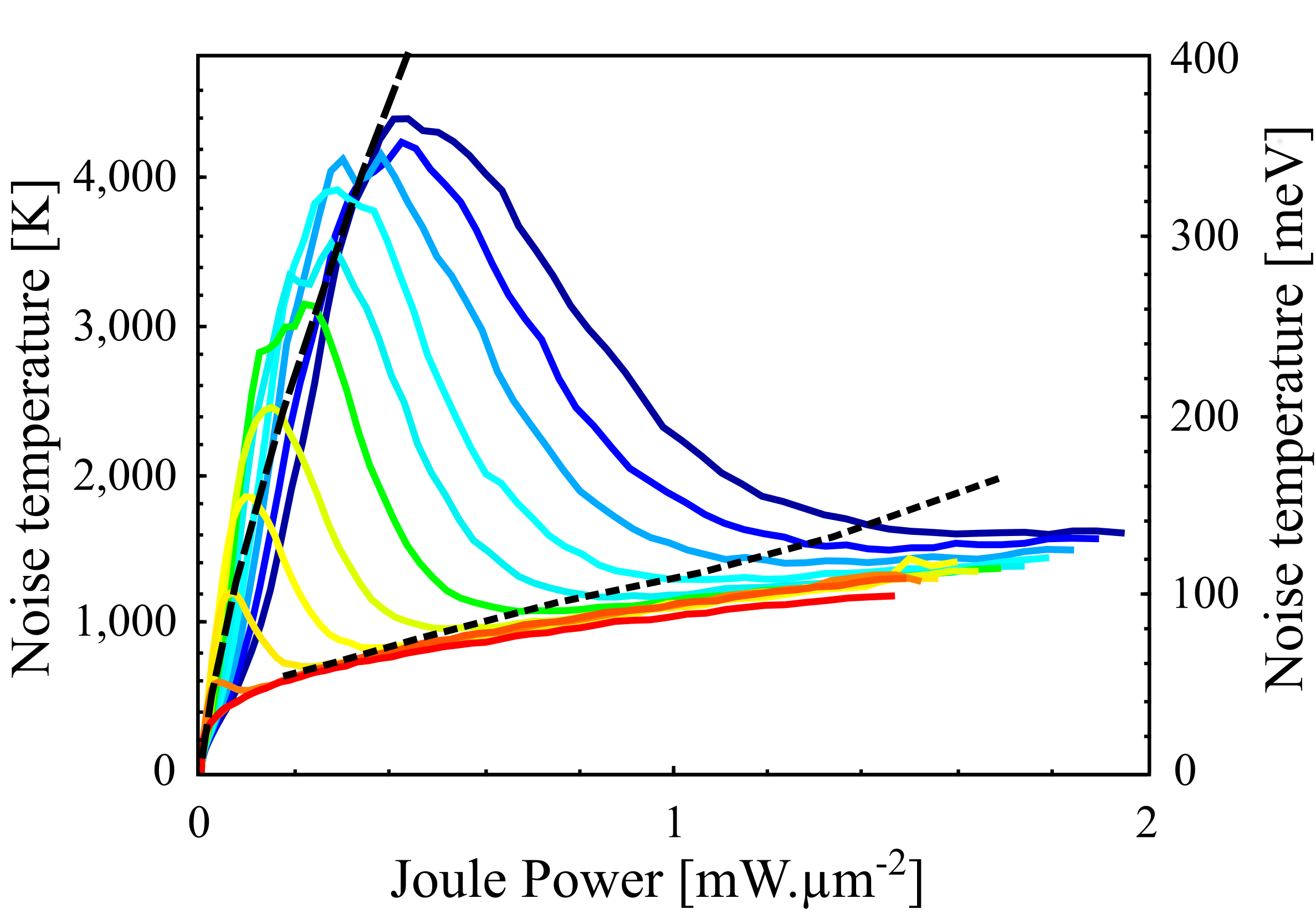}}
\caption{Hyperbolic phonon polariton cooling of a bilayer graphene transistor.
The noise temperature $T_{N}$ shows a non-monotonic dependence in
the applied Joule power with a characteristic drop at a doping-dependent
threshold between a low-bias hot electron regime and a high-bias cold-electron
regime signaling the onset of a new and very efficient cooling mechanism:
the electroluminescence of mid-infrared hyperbolic phonon polaritons. The cooling is characterized by the heat conductance $G_{th}= P /T_e\gtrsim  P / T_N$ which switches from $50\;\si{kW m^{-2}K^{-1}}$ at low bias to
 $1.3\;\si{MW m^{-2}K^{-1}}$ at high bias. \added{Colors represent doping with respective carrier densities $-n=0$ (red), $0.18$ (tangerine), $0.35$ (orange), $0.53$ (amber), $0.70$ (yellow), $0.88$ (chartreuse), $1.05$ (green), $1.23$ (turquoise), $1.40$ (cyan), $1.58$ (cerulean), $1.75$ (blue), $1.93\times10^{12}\;\si{cm^{-2}}$ (dark blue)}. \added{Electroluminescent cooling starts when Zener tunneling begins. Zener tunneling threshold is represented as a dashed line and increases with doping because of Pauli blocking. Dotted line represents the Joule power for which electroluminescent cooling balances Joule heating due to Zener current, see section \ref{sec:ElectroCooling}.}}
\label{TvsPJ.fig1}
\end{figure}

Nevertheless, the absolute cooling power remains rather limited ($100\;\si{kW.cm^{-2}}$)
in the case of graphene supported on hexagonal Boron Nitride (hBN) due to the narrow acceptance spectral
band of HPhPs ($\sim30\si{meV}$ in hBN) compared to the broadband spectrum
of the thermal emission of hot electrons in the channel. This is very
similar to the well-known poor efficiency of a bulb lamp (defined
as the fraction of power emitted in the visible window) because of
the broadband emission of thermal fluctuating charges in the infrared.
This limitation is \replaced{overcame}{overcome} in light-emitting diodes (LEDs) that achieve
up to 90\% electroluminescence conversion efficiency using a non-equilibrium
distribution of electrons and holes that recombine across the band gap,
thereby funneling efficiently the electrical power into the relevant
visible window [\onlinecite{Santhanam2012prl,Santhanam2013apl,Xue2015apl,Liebendorfer2018xxx}].
Similarly, emission in the narrow spectral RS bands
where the substrate sustains hyperbolic modes can be targeted by driving
the transistor in the Zener-Klein (ZK) tunneling regime. The Zener current becomes
equivalent to an electrical pumping of non-equilibrium electron-hole
pairs that efficiently recombine in the hyperbolic modes of the RS
bands. This striking effect is readily seen in Figure~\ref{TvsPJ.fig1},
where the steady-state electronic temperature of a graphene transistor
is monitored (by means of noise thermometry [\onlinecite{Betz2012prl,Betz2013nphys,Laitinen2014prb,Brunel2015jpcm,Crossno2016science,Yang2018nnano,Yang2018prl}]) as a function of injected
Joule power \footnote{\added{Note that noise temperature is larger than electronic temperature under high current driving, a distinction that we do not consider in this review.}}. The outstanding observation is a \emph{drop} of the temperature
for \emph{increasing} injected power above a certain threshold that
corresponds to the ignition of HPhP electroluminescence in the hyperbolic hBN substrate.

In this review, we show - based on both experimental results and theoretical
proposals - that radiative heat transfer can be considerably enhanced
to become a major cooling pathway by either engineering of optical
impedance matching between the transistor channel and a hyperbolic
substrate or by operating the device in a Zener-like regime where
electrical pumping of electron-hole pairs funnels efficiently the injected power into the propagating hyperbolic bands of the substrate.
The discussion is based on the emblematic graphene on hBN transistors but can easily be extended to other prospective active material/hyperbolic substrate couples
where superPlanckian radiative cooling could yield a major improvement.
The paper is organized as follows : in a first section we briefly
introduce hyperbolic materials and the concepts of thermal coupling
between a 2D electronic channel and a bulk hyperbolic substrate; then we investigate
the case of non-equilibrium situations and the electroluminescence
of HPhPs. \replaced{Finally}{Next}, we present experimental data for a bilayer graphene on hBN
transistor where the electronic temperature is monitored through noise
thermometry.

\section{Coupling between graphene and hyperbolic materials}

\label{hyperbolic-materials}

Computing the thermal radiation of an electronic channel in the electromagnetic modes requires the knowledge of the electromagnetic response of the dielectric encapsulating medium which is provided by the dielectric permittivity.
For an isotropic polar dielectric, the optical phonons couple to light and
provide an effective dielectric response of the form
\begin{equation}
\epsilon(\omega)=\epsilon_{\infty}\left(\frac{\omega_{LO}^{2}-\omega^{2}-j\omega\gamma}{\omega_{TO}^{2}-\omega^{2}-j\omega\gamma}\right), \label{eq:Dielectric}
\end{equation}
 where $\omega_{LO}$ is the longitudinal optical frequency, $\omega_{TO}$
is the transverse optical frequency and $\gamma$ is the decay rate
of the optical phonon resonance.
For high-quality materials, \replaced{$\mathrm{Re}(\epsilon(\omega))$ is negative}{$\mathrm{Re}(\epsilon(\omega))<0$} in
the range $[\hbar\omega_{TO},\hbar\omega_{LO}]$, as a consequence,
the material is opaque in this frequency range called Reststrahlen
band (RS) usually in the mid-infrared ($[170,200]\si{meV}$ for hBN).
A RS band is characterized by its central frequency $\Omega=\left(\omega_{LO}+\omega_{TO}\right)/2$
and \added{its} width $\Delta\omega=\omega_{LO}-\omega_{TO}$.  \deleted{For most materials considered in
this review, $\frac{\Delta\omega}{\gamma}\sim10-100$, consequently,
for the sake of simplicity, we will consider the limit $\gamma\rightarrow0^{+}$.}

An anisotropic dielectric material posesses a tensorial dielectric response. For uniaxial crystals, a class \replaced{including}{comprising} most dielectric
2D materials, in-plane (transverse $t$) and out-of-plane ($z$-axis) optical phonon energies are sufficiently separated to give rise to a splitting of the Reststrahlen band into two bands. As an example, hBN possesses two well separated bands corresponding to out-of-plane OPs (type-I band, [97,103]\si{meV}), and in-plane OPs (type-II band, [170,200]\si{meV}). The energy and width of hBN RS band II makes it a much more efficient heat conduction channel, therefore, by sake of simplicity, we shall focus specifically on this band in this review \added{(with $\Omega \equiv \Omega_{II}$)}. \footnote{A discussion of hyperbolicity type can be found in Reference [\onlinecite{Caldwell2014ncomm}].}
As a consequence, for layered 2D heterostructures, the dispersion of electromagnetic extraordinary waves takes the form :
\begin{equation}
\frac{k_{z}^{2}}{\epsilon_{t}}+\frac{k_{t}^{2}}{\epsilon_{z}}=\frac{\omega^{2}}{c^{2}} \label{eq:Dispersion}
\end{equation}
and, within a type II RS band, $\epsilon_{t}<0<\epsilon_{z}$ which
means that the dispersion relationship is hyperbolic. Note that this
occurs only for materials with low-loss optical phonon modes $\Delta\omega/\gamma\gg1$;
a category named hyperbolic phonon materials. The important consequence
is that $k_{t}>\omega/c$ becomes possible within the dispersion relationship,
which opens up propagative modes with high momenta that can help radiate energy.

The efficiency of hyperbolic radiative cooling depends critically
on two parameters: (i) the ability of the dielectric material to
efficiently couple to the current fluctuations of the metal, and (ii)
the propagation depth of this radiation within the hyperbolic dielectric.
These two aspects are best understood within the framework of transmission
line theory \footnote{This approach is possible thanks to the spatial Fourier decomposition
of eigenmodes allowed by in-plane translational invariance of the
structures considered.}: Considering a 2D conductor with random \footnote{We only consider longitudinal current fluctuations and do not consider the transverse current fluctuations because they do not couple to
the extraordinary wave which is TM polarized.} current density fluctuations $j(\omega,{\bf q}),$  where ${\bf q}$
is the in-plane wavector, this source interacts with the TM electromagnetic
modes of the substrate of optical impedance $Z(\omega,{\bf q})$
and the metal conductivity $\sigma(\omega,{\bf q})$ in parallel.
As a consequence, the power cast by this source in \replaced{its}{the} electromagnetic
environment  is ${\rm d}P=\frac{1}{2}\frac{\mathrm{Re}\left(Z^{-1}\right)}{\left|Z^{-1}+\sigma\right|^{2}}|j|^{2}.$
 Note that the optical impedance of the dielectric environment
\replaced{$Z(\omega,{\bf q})=Z_{xy}=|\mathcal{E}_{x}|/|\mathcal{H}_{y}|$}{$Z(\omega,{\bf q})=Z_{xy}=|E_{x}|/|H_{y}|$}  is defined as the ratio
between the longitudinal electric field collinear to the currents in
the material, and the transverse magnetic field of the extraordinary
wave. It fully determines the electromagnetic response of the environment
as seen from the channel's perspective. In terms of optics, the power
${\rm d}P$ can be interpreted as the average flux of the Poynting
vector through the channel/dielectric interface.

According to the fluctuation-dissipation theorem, for a single band,
the surface density of the current fluctuations in a material of
conductivity $\sigma$ at temperature $T$ reads
\begin{equation}
\overline{|j|^{2}}=\frac{1}{(2\pi)^{3}}4\mathrm{Re}\left(\sigma\right)~\hbar\omega\; n_{ph}(T,\omega)\,\mathrm{d}^{2}q\mathrm{d}\omega, \label{eq:CurrFluct}
\end{equation}
where \replaced{$n_{ph}(T,\omega)=\left[\exp(\hbar\omega/k_{\mathrm{B}}T)-1\right]^{-1}$}{$n_{ph}(T,\omega)=(\exp(\hbar\omega/k_{\mathrm{B}}T)-1)^{-1}$}
is the Bose occupation number. In our experiment, the hBN layer remains cold due to the presence of the gold backgate electrode. Therefore, we can neglect the back thermal flux from the hBN to the channel. The total radiative power is then \footnote{taking into account summation over positive and negative frequencies}
\begin{equation}
P_{\mathrm{rad}}=\frac{1}{(2\pi)^{2}}\int_{0}^{\infty}\mathrm{d}\omega\,\int_{0}^{\infty}q\mathrm{d}q\,\mathcal{M}(\omega,q)\,\hbar\omega\; n_{ph}(T,\omega),\label{eq:Prad-1}
\end{equation}
where $0\leq \mathcal{M}\leq1$ is the impedance matching \replaced{factor}{coefficient} :
\begin{equation}
\mathcal{M}(\omega,q)=\frac{4\mathrm{Re}\left(Z^{-1}\right)\,\mathrm{Re}(\sigma)}{|Z^{-1}+\sigma|^{2}}.\label{eq:MatchingFactor-1}
\end{equation}

The matching factor (\ref{eq:MatchingFactor-1}) restricts the phase
space available for radiative heat exchange. Indeed the optical impedance
reads $Z=\frac{Z_{0}}{\epsilon_{t}}\frac{k_{z}}{k_{0}}$ , where
$k_{0}=\omega/c,$ and $Z_{0}=377\;\added{\mathrm{Ohms}}$ is the vacuum impedance.
\begin{figure}[ht]
\centerline{\includegraphics[width=10cm]{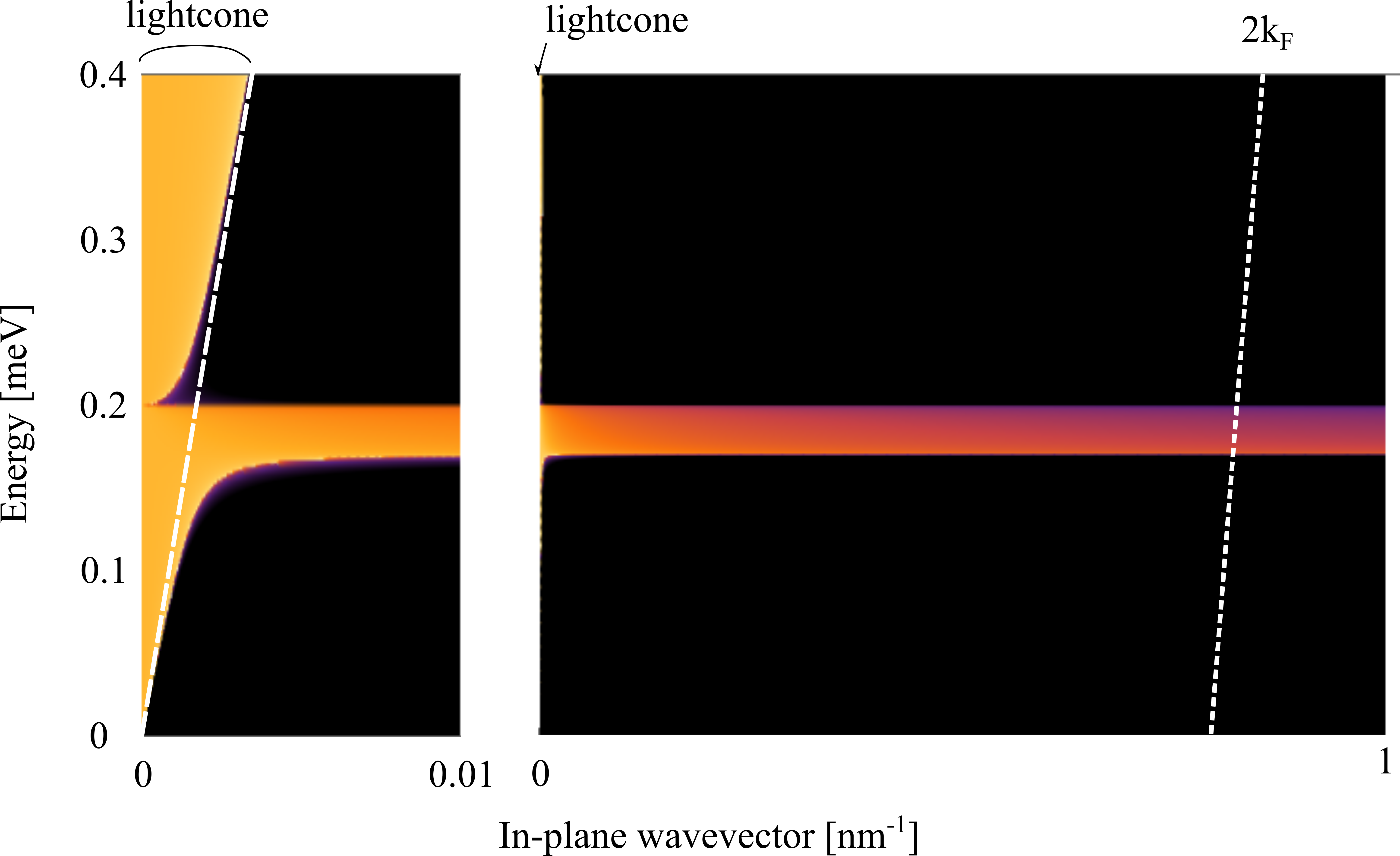}} \caption{Real part of the optical admittance ($Z^{-1}$) of hBN (omitting RS band
I) on two in-plane wavevector scales to illustrate the far field and
HPhP extension of radiation phase space. As discussed in section \ref{ssec-thermalhypcool}, the impedance matching window in the RS band is upper bounded by at least $2k_F$. }
\label{fig-ExempleRS}
\end{figure}
When the photon energy is outside the RS band, $k_{z}$ becomes imaginary
when $q>k_{0} \sqrt{\varepsilon_{t}},$ and thus $\mathcal{M}=0$.
As a consequence, radiation is limited to the very narrow lightcone
as represented on Figure \ref{fig-ExempleRS}  and the emitted power is very small. For monolayer graphene at $2000~\si{K}$ in the vacuum, it represents a radiated power $\bar{\mathcal{M}} \sigma_{SB} T^4\simeq 1.5~\si{W cm^{-2}}$, where $\sigma_{SB}=5.67\;10^{-8}\;\si{W m^{-2} K^{-4}}$ is the Stefan-Boltzmann constant, and $\bar{\mathcal{M}}\sim 2.3\%$ is the average far-field emissivity of graphene [\onlinecite{Nair2008science}]. This corresponds to a characteristic thermal conductance $\frac{P}{T} \simeq 10\, \si{W m^{-2} K^{-1}}$.   In contrast, within the RS band, $k_{z}$ remains real, and hence $\mathcal{M}$
remains sizeable for much larger $q$ vectors, enhancing considerably the emission efficiency.

At this point we distinguish two possible causes for current fluctuations:
(a) thermal fluctuations of charges (black-body radiation) and (b) non-equilibrium fluctuations
of charges (electroluminescence). The next two sections will address each of them.

\subsection{Thermal hyperbolic cooling}
\label{ssec-thermalhypcool}
\subsubsection{Matching Factor}

Let us first consider thermal emission by a single band. Since RS bands are narrow $\Delta\omega\ll\Omega$, \added{the photon population is nearly constant over a RS band and given by the occupancy at mid-band $n_{ph}(T,\omega)\simeq n_{ph}(T,\Omega)$. The radiated power (\ref{eq:Prad-1}) in the RS band energy then reads}
\begin{equation}
P_{\mathrm{rad}}=A_{F} k_{F}^{2}\hbar\Omega\; n_{ph}(T,\Omega)\Delta\omega\label{eq:Prad-1-1}\; ,
\end{equation}
with
\begin{equation}
A_{F}=\frac{1}{(2\pi)^{2}\Delta\omega}\int_{\omega_{_{TO}}}^{\omega_{LO}}\mathrm{d}\omega\,\int_{0}^{\infty}x\mathrm{d}x \mathcal{M}(\omega,k_Fx),
\end{equation}
where $k_{F}$ is the Fermi wavevector. $A_{F}$ can be interpreted as an average impedance matching factor over the RS band. For a metal with parabolic dispersion, $A_F \pi \hbar\Omega n_{ph}(T,\Omega)\Delta\omega$ also corresponds to the average power radiated per electron in the conduction band. Let us note that in the degenerate case, integral over wavevector (\ref{eq:Prad-1-1}) is bounded in a window due to a cut-off defined by Pauli blocking.

Since most optical modes lie out of the lightcone
\replaced{$k_{F}\gg \sqrt{\varepsilon_t}\Omega/c,$}{$k_{F}\gg k_{0}\sqrt{\varepsilon_t}=\sqrt{\varepsilon_t}\Omega/c,$} the optical impedance is well approximated
by $Z\simeq\left(\frac{Z_{0}}{\sqrt{\varepsilon_{\perp}^{\infty}\varepsilon_{z}^{\infty}}}\sqrt{\frac{\omega_{\mathrm{LO}}^{2}-\omega^{2}}{\omega^{2}-\omega_{\mathrm{TO}}^{2}}}\frac{q}{k_{0}}\right)$ \added{in the infinite dielectric-thickness limit }\footnote{\added{The finite thickness optical impedance can be found in Reference [\onlinecite{Yang2018nnano}]}}.
Regarding the 2D electronic channel response, due to the significant extension of the integration window in the first Brillouin zone, the nonlocal conductivity is needed.
We rely on the linear response theory for the case of non-interacting electrons in a single parabolic band:
\begin{equation}
\sigma(\omega,q)=iG_{0}\left(\frac{g_{s}g_{v}/2}{(2\pi)^{2}}\frac{\hbar\omega}{q^{2}}\right)\int d{\bf k}\ \frac{f_{{\bf k}}-f_{{\bf k}+{\bf q}}}{\hbar\omega+E_{{\bf k}}-E_{{\bf k}+{\bf q}}+i\hbar\eta}\left|\left\langle \chi_{{\bf k}}|\chi_{{\bf k}+{\bf q}}\right\rangle \right|^{2},
\end{equation}
 where $G_{0}=2e^{2}/h=77\ \si{\mu S}$ is the quantum of conductance,
$g_{s}=2$ and $g_{v}=2$ are the spin and valley degeneracy,
$f_{{\bf k}}=\frac{1}{\exp(E_{{\bf k}}-\mu)/k_B T+1}$ is the Fermi-Dirac
occupation factor for state of wavector ${\bf k}$, $E_{{\bf k}}$
is the band energy dispersion, $\mu$ is the chemical potential, $\eta$
is the effective relaxation rate of electrons, and $\left|\chi_{{\bf k}}\right\rangle $
is the periodic part of the Bloch wavefunction.

By sake of simplicity, we make the following assumptions:
\begin{enumerate}
\item Isotropic parabolic dispersion: Except for monolayer graphene, isotropic parabolic dispersion covers most naturally occurring cases. The specific case of graphene has been dealt in detail in the literature [\onlinecite{Principi2017prl}]. \item Unitary Bloch wave projection: Due to the small extent of the Fermi sea compared to the Brillouin zone, we can approximate $\left|\left\langle \chi_{{\bf k}}|\chi_{{\bf k}+{\bf q}}\right\rangle \right|^{2}\sim1.$
\end{enumerate}
The computation of the nonlocal conductivity is given in appendix. In
the degenerate case, one has $\sigma(\omega,q)=G_{0}\frac{g_{s}g_{v}}{4\pi}S(\mu/\hbar\omega,q/k_{F})$,
where $S$ is a simple geometric function. Consequently,
the
prefactor $A_{F}$ is a function of the reduced doping $\mu_r=\mu/\hbar\Omega_{II}$
and of an effective light-matter coupling factor that reads
\begin{equation}
\alpha_{HPhP}=\mathrm{Re}(Z(\Omega,q_c))\mathrm{Re}(\sigma(\Omega,q_c))=\alpha_{FS}\ \frac{g_{s}g_{v}}{2\pi \sqrt{\varepsilon_{\perp}^{\infty}\varepsilon_{z}^{\infty}}}\frac{k_F}{k_0}\; ,
\end{equation}
 where $\alpha_{FS}=G_{0}Z_{0}/4=1/137$ is the fine-structure constant, and $q_c=k_F/\sqrt{\mu_r}=\sqrt{2m^*\Omega/\hbar}$, where $m^*$ is the effective mass of the parabolic electronic band. The doping dependence of $\alpha_{HPhP}$ reflects the ability of the \replaced{channel's electrons}{electrons in the channel} to couple to the HPhP modes of the substrate.

\begin{figure}
\begin{centering}
\includegraphics[width=10cm]{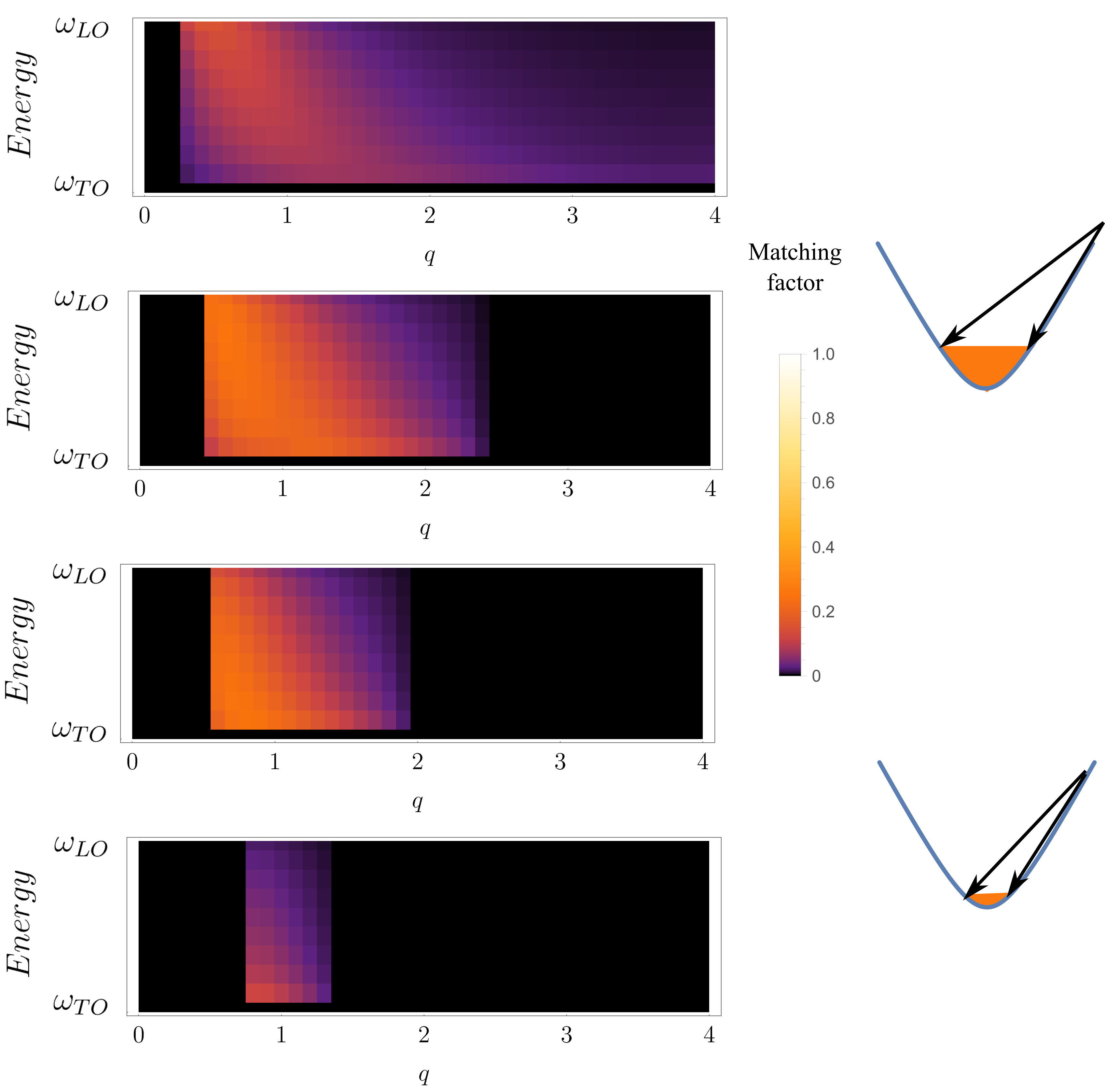}
\par\end{centering}
\caption{Impedance matching factors $\mathcal{M}$ (color scale) in RS band II of hBN (single band approximation for bilayer graphene with $m^*=0.03m_0$) for increasing doping (from bottom to top, $\mu_r=\mu/\hbar\Omega_{II}=\{0.1,0.5,1.0,5.0\}$). Wavevector is given in units of $q_c=\sqrt{2m^*\Omega_{II}/\hbar}\simeq4.4\, 10^8\,\si{m^{-1}}$. The emission window increases with doping up to $\sim 2k_F$ at large doping, while the average matching is reduced over the RS band.}
\label{fig:MatchingFactorNullTemp}
\end{figure}

Figure \ref{fig:MatchingFactorNullTemp} represents the matching factor taking into account only the conduction band of bilayer graphene for increasing doping. The allowed transitions in the degenerate regime restrict the HPhP radiation window in the wavevector range $[(\sqrt{1+\mu_r^{-1}}-1) k_F,\sqrt{1+\mu_r^{-1}}+1) k_F]$, which is centered on $q_c=k_F/\sqrt{\mu_r}=\sqrt{2m^* \hbar \Omega_{II}}/\hbar \simeq4.43\, 10^8\ \si{m^{-1}}$.

For semimetals (like few layer graphene), the wavevector window for impedance matching is larger and extends over $[0,\sqrt{2}q_c]$ in the degenerate case, due to the allowed interband transitions in the thermal regime. For  moderately doped graphene, $\mu<\hbar\Omega_{II}/2$, the weak electron-photon coupling $\alpha_{FS}$ is balanced by the characteristic electronic transition
wavevector with respect to optical wavevector $q_c/k_{0}\sim 460$, so that $\alpha_{HPhP}\sim 1$.
We conclude that graphene is \emph{naturally matching the naturally hyperbolic material} hBN.
The number of layers of graphene itself impacts obviously the radiated power due to different electronic dispersion. As a rule of thumb, we estimate a radiative cooling power of $1.6\;\mathrm{GWm^{-2}}$ at a temperature of $2000\;\mathrm{K}$ which translates into a thermal conductance $0.8\;\mathrm{MWm^{-2}K^{-1}}$ intermediate between the experimental values in Figure \ref{TvsPJ.fig1}. Finally, note that experimental observations reveal that HPhP cooling is suppressed when graphene is submitted to a quantizing magnetic field  [\onlinecite{Yang2018prl}] because electrons localize in small cyclotron orbits forbidding wavevector matching with HPhPs.

\subsubsection{\replaced{Propagation depth}{Attenuation Length} of the HPhPs}

The above evaluation of the radiated power doesn't provide information on where heat is dissipated. To obtain an efficient heat release, the propagation \replaced{depth}{length} of HPhPs must be large enough to release heat remotely in a large volume \replaced{and}{to} avoid hot-phonon effects. The characteristic \replaced{propagation depth}{attenuation Length} of the HPhPs is given by $l=(\mathrm{Im}(K_{z}))^{-1}.$ For large in-plane k-vector compared
to $k_{0}$, the characteristic wavevector in a RS band is $K_{z}\simeq\sqrt{-\frac{\varepsilon_{\perp}}{\varepsilon_{z}}}q.$
At center frequency $\Omega$, field propagation is \replaced{maximal}{minimal} and reaches
\begin{equation}
l\simeq\frac{\Delta\omega}{\gamma}\sqrt{\frac{\varepsilon_{z}^{\infty}}{\varepsilon_{\perp}^{\infty}}}\frac{1}{q}.
\end{equation}
In the case of bilayer graphene on hBN, the characteristic wavector of near-field radiative heat exchange is $q=k_{F}\sim 200k_{0},$ for which the \replaced{propagation depth}{attenuation Length} reaches 240 \si{nm}. Note that the presence of a nearby metallic back gate, with a low optical impedance, will reflect HPhPs and eventually lead to the reabsorption of the emitted power after one round trip. This situation occurs in thin hBN samples such as in Reference [\onlinecite{Yang2018nnano}]\deleted{ (main text)}. This is avoided in the back-gated device analyzed below where the hBN thickness ($200\;\si{nm}$) exceeds $\delta/2\simeq120\;\si{nm}$, so that a full HPhP cooling can be observed.

\subsection{Electroluminescence in a Hyperbolic Material}

Electroluminscence is associated with interband optical conductivity  occurring when electron-hole pairs are electrically pumped from the valence to the conduction band, a situation which occurs spontaneously in high mobility graphene when a large bias is applied [\onlinecite{Yang2018nnano}]. The electroluminescence results from the interband recombination of electron-hole pairs. Whereas classical electroluminescence is limited to vertical transitions, this emission process can also be considerably boosted when high-$\vec{k}$ propagating modes are available in the dielectric. In this section we focus on the excess radiation power related to electroluminescence.

The theoretical treatment of electroluminescence is closely related to near-field thermal emission, except that out-of-equilibrium current fluctuations are given by the nonlocal van Roosbroeck-Shockley relation for interband transitions  [\onlinecite{Wurfel1982jpc,Rosencher2002cambridge,Greffet2019prx}]:

\begin{equation}
\overline{|j_{\mathrm{inter}}|^{2}}=\frac{1}{(2\pi)^{3}}4\mathrm{Re}\,(\sigma_{inter})~\hbar\omega\; n_{ph}(T,\omega)\,\mathrm{d}^{2}q\mathrm{d}\omega, \label{eq:CurrFluct2}
\end{equation}
where \replaced{$n_{ph}(T,\omega)=\left[\exp(\hbar\omega-\mu_{ph})/k_{\mathrm{B}}T-1\right]^{-1}$}{$n_{ph}(T,\omega)=(\exp(\hbar\omega-\mu_{ph})/k_{\mathrm{B}}T-1)^{-1}$}
with $\mu_{ph}=\mu_{\lambda}-\mu_{\lambda'}$ is the photon chemical
potential associated to the difference in chemical potential between
the two bands of interest (denoted as $\lambda$ and $\lambda'$). Note that this approximation assumes an ultra-fast electron-electron thermalization, and the existence of two pseudo Fermi distributions in the valence and conduction bands. The interband nonlocal conductivity, $\sigma_{inter}(\omega,q)$, is evaluated on the interband transitions between the bands of interest. The total radiated power is the sum of
power radiated by all intraband and interband transitions. Focusing
on the case of a pair of valence and conduction bands, the fraction
of radiated power cast by electroluminescence is deduced from
Eqs.(\ref{eq:Prad-1}) and (\ref{eq:MatchingFactor-1}) using $j_{\mathrm{inter}}$
as the source. Note that, in this case, the impedance matching coefficients
read
\begin{equation}
\mathcal{M}(\omega,q)=\frac{4\mathrm{Re}\left(Z^{-1}\right)\,\mathrm{Re}\left(\sigma_{\mathrm{inter}}\right)}{|Z^{-1}+\sigma_{\mathrm{inter}}+\sigma_{\mathrm{intra}}|^{2}},\label{eq:MatchingFactor-1-1}
\end{equation}
where $\sigma_{intra}$ is the intraband conductivity.

With gaped semiconductors, electroluminescence becomes important
when the difference in conduction and valence chemical potential reaches
the bandgap energy $\mu_{c}-\mu_{v}\lesssim E_{g}$. In contrast to
this well-known configuration, graphene has no bandgap, however,
the threshold energy is now the RS band energy $\mu_{c}-\mu_{v}\lesssim\hbar\Omega_{II}.$
For unbiased graphene, out-of-equilibrium electronic distribution
can be transiently created by optical pumping and have been observed
to relax on sub-ps timescales by optical pump-probe techniques [\onlinecite{Ong20192DM,Mak2014apl,Tielrooij2018nnano,Malic2017andp}]. In the next section, we show that electrical pumping of charges occurs similarly in high mobility graphene under large bias and leads to a favorable situation for HPhP electroluminescence cooling.

\section{Hyperbolic phonon polariton electroluminescence}
\label{sec:ElectroCooling}

\begin{figure}[ht]
\centerline{\includegraphics[width=16cm]{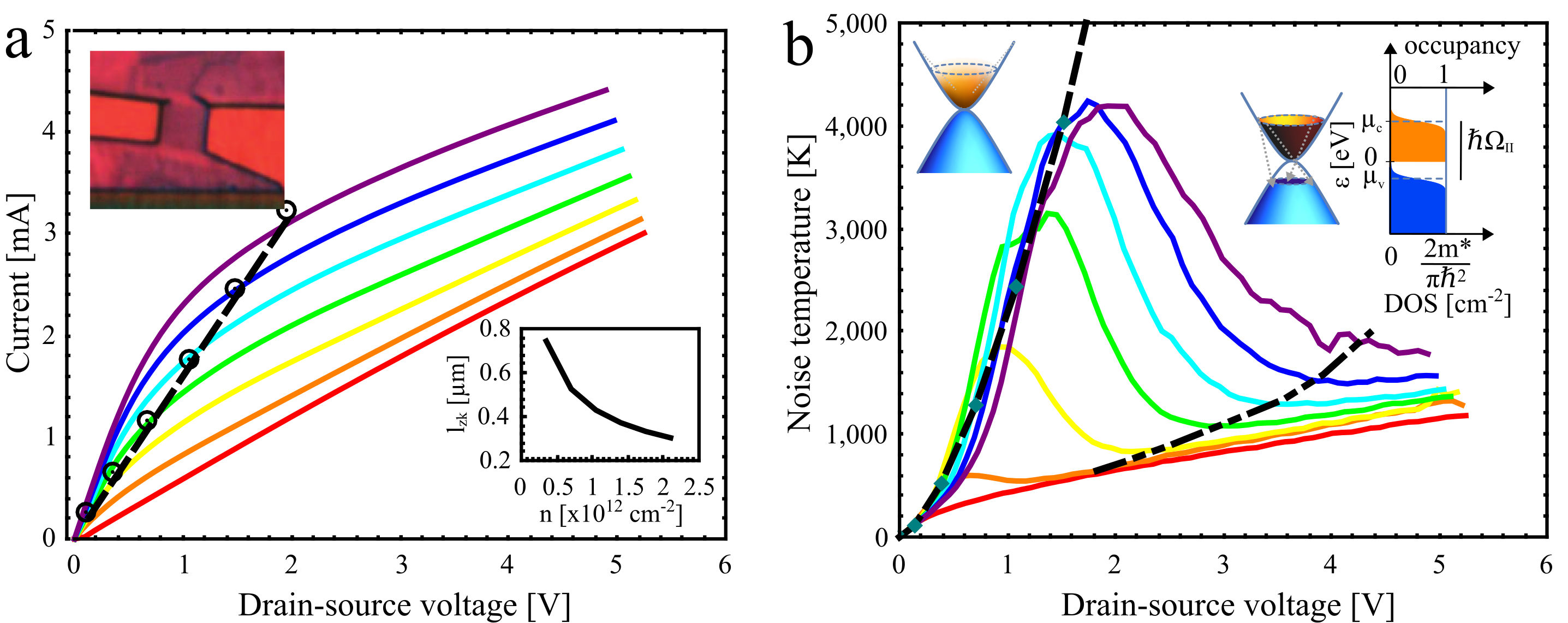}}
\caption{Transport and noise of a hBN-supported graphene ZKT transistor.
Panel a) current saturation and Zener-Klein regimes in a high-mobility
bilayer graphene flake of dimensions $L\times W=3.6\times3\;\si{\mu m}$.
Sample (upper inset) is deposited on a $200\;\si{nm}$ thick hBN
flake acting both as a bottom gate dielectric (capacitance $C_{g}=0.14\; \si{mF\, m^{-2}}$)
and hyperbolic phonon polariton radiator. Carrier density is tuned
in the hole doping range \added{$-n=0$ (red), $0.35$ (orange), $0.70$ (yellow), $1.05$ (green), $1.40$ (cyan), $1.75$ (blue), $2.10\times10^{12}\;\si{cm^{-2}}$ (purple)}. At low
bias intraband current dominates up to a threshold field $\mathcal{E}_{sat}=v_{sat}/\mu$
set by the saturation velocity $v_{sat}\lesssim 3\times10^{5}\;\si{m\, s^{-1}}$
and mobility $\mu\gtrsim1\;\si{m^{2}V^{-1}s^{-1}}$. At high bias,
above a Pauli unblocking voltage $V_{zk}$ (see text), intraband current
is fully saturated and an additional current arising from interband
Zener-Klein tunneling with a constant conductance $G_{zk}=0.45\pm0.05\;\si{mS}$
sets in. The boundary between intraband and interband transport regimes
follows theoretical expectation $I_{sat}\simeq\frac{2}{\beta_{zk}}G_{zk}V_{zk}$ (black
dotted line with the squares corresponding to theoretical values of
current). Panel b) Noise temperature $T_{N}=S_{I}/\left(4G_{ds}k_{B}\right)$
deduced from the microwave excess current noise $S_{I}$ and the differential
conductance $G_{ds}$. Noise reveals a striking difference between
the intraband and interband regimes with a superlinear bias dependence
characteristic of \textquotedbl hot\textquotedbl{} electrons (dotted
black line) in the former, dropping toward a linear dependence characteristic
of \textquotedbl cold\textquotedbl{} electron regime in the latter.
The theoretical value for the threshold between electroluminescent
cooling and heating is sketched as the dashed \deleted{blue} line. As sketched
in the inset, the tunneling process creates an out-of-equilibrium
electron-hole distribution which relaxes under HPhP emission.}
\label{fig:Transport}
\end{figure}
\label{electroluminescence}

To illustrate HPhP cooling we analyze here one of the
graphene ZKT transistors investigated in Reference {[}\onlinecite{Yang2018nnano}{]}
(Figure S4 of the Supplementary Material). It is a high-mobility bilayer
graphene flake, of dimensions $L\times W=3.6\times3\;\si{\mu m}$,
deposited on a $200\;\si{nm}$-thick hBN crystal (inset of Figure \ref{fig:Transport}-a).
A global metallic back gate is used to tune electron density in the
range $-n=[0,2.1\times10^{12}]\;\si{cm^{-2}}$. Figure \ref{fig:Transport}-a
depicts the transport properties of the ZKT transistor which is characterized
by an initial increase of the intraband current up to a doping-dependent
saturation value $I_{sat}$ followed by an interband contribution
with a doping-independent differential conductance $G_{zk}\simeq0.45\;\si{mS}$.
Figure \ref{fig:Transport}-b shows the noise temperature characteristics.

\subsection{Intraband current saturation}

At moderate bias in high mobility devices, we observe a current partial
saturation {[}\onlinecite{Meric2008nnano}{]} which is well described by the semi-empirical expression of the conductivity
$g_{ds}=G_{ds}\times L/W=|n|e\mu/[1+(\mathcal{E}/\mathcal{E}_{sat})^{2}]$ {[}\onlinecite{Yang2018nnano}{]}.
Fits \replaced{of}{from} the differential conductivity allows obtaining mobility
$\mu\gtrsim1\;\si{m^{2}V^{-1}s^{-1}}$ and a doping-dependent saturation electric field $\mathcal{E}_{sat}\lesssim0.3\;\si{V\, \mu m^{-1}}$. Mobility
is large enough for intraband current saturation to be achieved at
low to moderate voltage $V_{ds}\lesssim1\;\si{V}$, leaving a broad experimental window for inter-band transport. The current
saturation characteristic energy $E_{sat}$  is deduced from the saturation electric
field ${\cal E}_{sat}$ with $E_{sat}=\frac{hne\mu}{e(4k_F/\pi)}{\cal E}_{sat}$.  [\onlinecite{Yang2018nnano}]
At the low to moderate doping of our experiment, we remark that current saturation
is limited by the Fermi energy $E_{sat}=E_{F}$ [\onlinecite{Yang2018nnano}]\deleted{(not shown)}.

\subsection{Zener-Klein tunneling}

We recall here the properties
of the high-bias Zener-Klein regime introduced in Reference {[}\onlinecite{Yang2018nnano}{]}.
Assuming a uniform electric field, one can express the doping-independent
Zener-Klein resistance as the series addition of $L/l_{zk}$ incoherent
tunneling processes taking place over a length $l_{zk}$ with an elementary
conductance $G_{1}=\frac{1}{2}\beta_{zk}MG_{0}$, where $M=k_{F}W/\pi$
is the number of transverse electronic channels involved in electronic
conduction, and $\beta_{zk}\lesssim1$ is a smooth junction factor \footnote{$\beta_{zk}$ was labeled $\alpha_{zk}$ in reference [\onlinecite{Yang2018nnano}]}
(bilayer graphene [\onlinecite{Yang2018nnano}]). This leads
to an apparent Zener-Klein conductivity $\sigma_{zk}=\beta_{zk}G_{0}\times\frac{Ml_{zk}}{W}\propto\beta_{zk}k_{F}l_{zk}$,
because each electron-hole pair spontaneously created by Zener tunneling diffuses over a length $l_{zk}$ before recombining separately with conjugated charge carriers.  Experimentally, one can extract the product $\beta_{zk}l_{zk}\propto1/k_{F}$ from the constant differential conductance $G_{zk}\simeq0.45\;\si{mS}$. Zener-Klein tunneling
being controlled by Pauli blocking, its ignition is subject to a threshold field $\mathcal{E}_{zk}\simeq2|E_{F}|/el_{zk}$. If mobility is large
enough, the intraband current is saturated prior to the onset of Zener-Klein tunneling so that $J(\mathcal{E}_{zk})\simeq J_{sat}$. This defines the intraband/interband transport boundary as the line $I_{sat}=\frac{2}{\beta_{zk}}G_{zk}V_{zk}$,
highlighted as a black dotted line in Figure \ref{fig:Transport}-a.
Note in passing that saturation currents
do reach the intrinsic limit (symbols in the dotted line of  Figure \ref{fig:Transport}-a) in this sample. From the slope of the
line we extract $\beta_{zk}\simeq0.5$, and finally deduce $l_{zk}(|n|)$
(lower inset of Figure \ref{fig:Transport}-a). The $l_{zk}\propto1/\sqrt{|n|}$
dependence, indicates that the electron-hole diffusion length is proportional
to the average distance between charges. Remarkably enough, transport
does not show any indication of thermal degradation at high bias in
spite of the large Joule power $P=200\;\si{kW cm^{-2}}$, which
is three orders of magnitude larger than the limit for thermal degradation
in MOSFETS [\onlinecite{Pop2010nres}]. To get a deeper insight
on the physical origin of this hyper-cooling we rely on the noise
analysis below.

\subsection{Hyperbolic phonon polariton cooling}

As explained in the
experimental section below, a Johnson-Nyquist noise temperature $T_{N}=S_{I}(\omega)/4k_{B}G_{ds}$
can be deduced from the GHz current noise $S_{I}$ and differential
conductance $G_{ds}$. Its bias voltage dependence $T_{N}(V_{ds})$
in Figure \ref{fig:Transport}-b reveals very contrasted behaviors for
intraband and interband transport. The former corresponds to a  a superlinear bias dependence $T_{N}\propto V_{ds}^{1.5}$
at low bias [\onlinecite{Yang2018nnano}] and  a quasi-linear dependence $T_{N}\propto V_{ds}$ at high bias. Remarkably the transition occurs at the Zener-Klein tunneling
onset defined above according to Pauli blocking (black dotted line),
and gives rise to a dramatic drop of the noise temperature (or a noise  plateau \replaced{for graphene transistors with thin hBN dielectric}{in thin hBN transistors} {[}\onlinecite{Yang2018nnano}{]}).
The neutrality regime provides a valuable insight on the cooling mechanism, by featuring a threshold at $V_{ds}=0.2\;\si{V}$, which is invariantly observed in graphene allotropes (single-layer, bilayer and trilayer graphene), pointing to a phonon activation mechanism with an energy $\simeq0.2\;\si{eV}$ matching the type-II HPhP band of hBN [\onlinecite{Yang2018nnano}].
This activation energy corresponds to the ``optical gap'' $\hbar\Omega_{II}$ expected for HPhP electroluminescence.

If the low-bias regime can be interpretaed in the framework of superplanck HPhP black-body emission, the high-bias regime cannot (for details see Reference [\onlinecite{Yang2018nnano}]). In particular, superplankian emission cannot account for the drop of temperature in increasing Joule power. An out-of-equilibrium emission of HPhPs by hot electrons in the inter-band regime is needed.

\subsection{Electroluminescence hyperbolic phonon polariton cooling}

A parallel with LED is helpful to understand why electroluminescence is associated to a cooling of the electron gas. In LEDs, electron-hole pairs are injected from the electrodes and recombine in the active region. The energy per injected electron-hole pair  $E_{inj}$ is slightly
larger than the effective bandgap at the junction, and consequently, radiative recombination energy $E_{g}<E_{inj}$ is insufficient to sink the injected energy (the situation is made even worse considering the competing nonradiative recombination). As a consequence, electroluminescence
is usually associated with a small heating of the device. However, it has been reported that, for high quality devices at subthreshold bias for which $E_{g}>E_{inj}$, the LED acts as a refrigerator as the missing energy $E_{g}-E_{inj}$ is provided by the thermal energy of the electron gas. To date, LED refrigerators are very inefficient with a record cooling power of only $8\;\si{pW}$ [\onlinecite{Santhanam2013apl}].  The graphene transistor presented in this paper operates in refrigerator mode at moderate bias where  $E_{inj}<\hbar\Omega_{II}$, therefore, the sudden drop in temperature at the Zener-Klein tunneling
threshold is a consequence of a \replaced{competition}{balance} between intraband heating,
and interband HPhP electroluminescent cooling. The cooling power reaches
$\sim10\si{mW}$, i.e., nine orders of magnitude larger than  in conventional LEDs. However, since the electron gas temperature is much larger than the phonon bath one, as a whole this graphene transistor is not a refrigerator and cooling power is not bounded by a Carnot
inequality, in contrast with LEDs where the device gets colder than
its environment.

The injected energy per electron-hole pair is the electrical work done by the charge during diffusion and reads $E_{inj}=P_{zk}/\dot{n}_{zk}$, where $P_{zk}=\sigma_{zk}({\cal{E}}-{\mathcal{E}}_{zk}){\cal{E}}$ is the Joule power associated with Zener tunneling processes, and $\dot{n}_{zk}=G_0 M  ({\cal{E}}-{\mathcal{E}}_{zk})/e$ is the rate of electron-hole pair creation by Zener-Klein tunneling. We deduce the injection energy per electron-hole pair to be $E_{inj}=\beta_{zk}l_{zk}e{\cal{E}}$. This allows obtaining the characteristic electric field for which injection energy balances HPhP emission energy $\hbar\Omega_{II}$ by electron-hole pair recombination.
This electric field ${\mathcal{E}}_{th}=\frac{\hbar\Omega_{II}}{\beta_{zk}l_{zk}e}$ defines the boundary between electroluminescent cooling and heating, it is indicated in Figure \ref{fig:Transport}-b (dashed line) and it matches well the voltage for which the minimal temperature is reached.

The chemical potential imbalance between the valence and conduction band can be estimated from the data in Figure \ref{TvsPJ.fig1}. For $-n=1.25\times10^{12}\;\si{cm^{-2}}$
doping, the maximum temperature $T_{N}\simeq3200~\si{K}$ is reached at ZK threshold for an injected Joule power $0.25~\si{mW.\mu m^{-2}}$, and drop to a minimal temperature $T_{N}\simeq1400~\si{K}$ for $1~\si{mW.\mu m^{-2}}$.
Assuming the cooling power variation with the electronic bath state is mainly governed by the photon occupation factor $P_{rad}\rightarrow n_{ph}$ (a legit assumption because of its singular behavior at the electroluminescence--superluminescence
threshold $\hbar\Omega_{II}=\mu_{c}-\mu_{v}$), we deduce that the
chemical potential imbalance $\mu_{c}-\mu_{v}\simeq 0.16\ \si{eV}\lesssim\hbar\omega_{TO}$
reaches the lower boundary of the hBN RS band II.
Below Zener-Klein electric field threshold, the electronic temperature results from a power balance between intraband processes only. At very high electric field, with Zener-tunneling and electroluminescence recombination increasing, the \deleted{state of the} conduction \deleted{band }and valence band reach a doping state which is mainly independent of gate-doping: $\mu_c-\mu_v \sim \hbar \Omega_{II}\gg (\mu_c+\mu_v)/2=\mu_0$, where $\mu_0$ is the Fermi energy at vanishing bias. As a consequence, the power balance tends to be dominated by Joule heating originating from ZK tunneling processes, which is independent of doping, and HPhP emission by both intraband thermal emission and interband electroluminescence defined by a similar electronic distribution. Therefore, noise temperatures obtained at finite doping tend to line up on the temperature obtained at null doping at large bias as observed on Figure \ref{TvsPJ.fig1}.
\deleted{The electronic temperature shifts from a fully intraband transition-driven power balance to a fully interband transition-driven power balance. For undoped graphene, the interband transitions dominate heat exchange at all biases, therefore, as expected observations at finite doping tend to line up on the null doping observation at large bias.}

\section{Conclusions and perspectives}

The above noise thermometry measurements in the Zener-Klein regime of a hBN-supported bilayer graphene transistor, supported by the theoretical analysis, demonstrate the superior cooling power of radiative HPhP electroluminescence. As shown in  Reference [\onlinecite{Yang2018nnano}] the mechanism is quite generic and consistently observed in single layer, bilayer and trilayer graphene. The role of hBN thickness is emphasized here with a stronger cooling effect, and a negative thermal conductance regime reported in the semi-infinite hBN substrate regime. Theoretical modeling provides useful keys to assess the difference between supported and encapsulated graphene, and constitutes a base to tackle more involved situations such as the rich moir\'e phases of hBN encapsulated graphene [\onlinecite{Bistritzer2011pnas,Cao2018nature,Yankowitz2019science}]. Hyperbolicity of the phonon modes adds to the  list of hBN qualities which include dielectric strength, compatibility with high graphene mobility [\onlinecite{Banszerus2015sciadv,Banszerus2016nl}], and
optical properties
[\onlinecite{Schue2016nscale,Schue2019prl}].

 HPhP radiative cooling is not limited to the hBN-graphene couple. It should also be observed with graphene on calcite [\onlinecite{Salihoglu2019jqsrt}] or mica [\onlinecite{Low2014small}]
which have similar hyperbolic Restrahlen bands.
Graphene on mica transistors have  a lower mobility possibly precluding the electroluminescent regime but still constitute  are valuable
route for high performance flexible electronics [\onlinecite{Hu2018as}]. Conductive materials such as cuprates or tetradymites (e.g. Bi$_{2}$Se$_{3-x}$Te$_{x}$)
also sustain hyperbolic modes {[}\onlinecite{Narimanov2015nphot}{]}. The case of tetradymites deserves special attention as HPhP energy lies in the near infrared to optical range [\onlinecite{Esslinger2014acsphot}],
which is larger than the bandgap. Their finite conductivity comes from topological surface states which can be used to realize field-effect transistors in the bulk depleted regime [\onlinecite{Inhofer2018prapp}]. Remarkably they combine, in one and the same material, the two properties valued in graphene-hBN heterostructures: a gapless Dirac Fermion transport
at the surface and hyperbolic HPhP modes in the bulk. Finally, and possibly most impactful for applications, superplanck\added{ian} HPhP cooling can be envisioned in silicon on insulator (SOI) technology [\onlinecite{Taur1997ieee}], provided that the $10\;\si{nm}$-thin active channel is nested in the near-field of a hyperbolic dielectric such as $\alpha$-sapphire or $\alpha$-quartz [\onlinecite{Winta2019prb}]. The inclusion of an HPhP cooling stage in MOSFET technology  would lift the thermal bottleneck at $\sim100\;\si{W\, cm^{-2}}$ reported in Reference [\onlinecite{Pop2010nres}].

\section{Experimental section}

\subsection{Sample}

Sample fabrication is described in Reference [\onlinecite{Yang2018nnano}]. The sample analyzed here is made of an as-exfoliated bilayer graphene
flake, of dimensions $L\times W=3.6\times3\;\si{\mu m}$, deposited
on a $200\;\si{nm}$-thick hBN crystal (inset of Figure \ref{fig:Transport}-a).
The use of a thick hBN dielectric (i.e. thicker than the HPhP \replaced{propagation depth}{damping
length}) avoids HPhP reflections at the metallic back gate and simplifies
analysis.  The sample is equipped with high-transparency
contacts, a global metallic backgate, and is embedded in a coplanar
wave guide (CPW) for high-frequency current noise measurement. DC
current and noise are measured at liquid helium temperature (4.2\si{K}),
in the hole doped regime $-n=0$--$2.1\times10^{12}\;\si{cm^{-2}}$,
where we obtain a low-field mobility $\mu\gtrsim1\;\si{m^{2}V^{-1}s^{-1}}$.

\subsection{Noise thermometry principles}

Noise thermometry principles, described in Reference [\onlinecite{Yang2018nnano}] (Supplementary), are introduced here. Electrons in a metal sustain thermal fluctuations leading to the well-known Johnson-Nyquist white current noise $S_{I}=4Gk_{B}T_{e}$
{[}\onlinecite{Nyquist1928prb}{]}, proportional to the sample conductance
$G$ and electronic temperature $T_{e}$. This Nyquist formula serves
as a primary thermometry and is used to estimate the out-of-equilibrium
electronic temperature at finite bias {[}\onlinecite{Betz2012prl,Betz2013nphys,Brunel2015jpcm,Laitinen2014prb,Yang2018nnano,Yang2018prl,Andersen2019science}{]},
when the heated electron gas decouples from the lattice and gets thermalized thanks to efficient electron-electron relaxation (rate $\sim10\;\si{fs}$). To account for self-heating in the Nyquist formula, one
substitutes $T_{e}$ by a noise temperature $T_{N}\simeq\langle T_{e}\rangle_{L,W}$
\footnote{This approximation is fair at low to moderate drift velocity but may
turn as an overestimate at large drift due to significant $\propto(v_{d}/v_{F})^{2}$
corrections}, and the linear conductance by the differential conductance $G_{ds}=\partial I_{ds}/\partial V_{ds}$,
taken at the noise measuring frequency.

Experimentally the finite currents entail an additional $1/f$-like
conductance noise, $S_{I}=\alpha_{H}I^{2}/Nf$ with $\alpha_{H}\sim 2\;10^{-3}$
is the Hooge constant and $N=|n|LW$ the total number of carriers
participating to transport [\onlinecite{Hooge1981rpp}],   obscuring thermal noise of interest at low frequency. Writing
thermal noise as $S_{I}=2e\mathcal{F}I$, where $\mathcal{F}\sim0.1$
is a typical thermal Fano factor, one estimates a $1/f$-corner frequency, $f_{c}\sim\frac{\alpha_{H}}{2\mathcal{F}}\times v_{d}/L$.
The large drift velocities $v_{d}\lesssim3\times10^{5}\;\si{m\, s^{-1}}$ of our high-mobility
graphene yields $f_{c}\lesssim1\;\si{GHz}$, whence the need to
measure thermal noise in the microwave band $f=1$--$10\;\si{GHz}$.

\label{ZKTtime}

\section{Supporting Information}

\subsection{Impedance matching in the degenerate case}

In this appendix, we compute a generic expression for the nonlocal conductivity of a single parabolic electronic band in the degenerate regime $k_B T \ll\mu$. We then compute the coupling factor $A_F$ in the case of HPhP thermal radiation within a RS band of energy $\hbar\Omega$ and width $\hbar \Delta\omega$.
\begin{equation}
A_{F}=\frac{1}{(2\pi)^{2}\Delta\omega}\int_{\omega_{_{mid}}-\Delta\omega/2}^{\omega_{_{mid}}+\Delta\omega/2}\mathrm{d}\omega\,\int_{0}^{\infty}x\mathrm{d}x \mathcal{M}(\omega,k_F x),
\end{equation}

In the \deleted{case }degenerate case,

\begin{equation}
\sigma(\omega,q)=iG_{0}\left(\frac{g_{s}g_{v}/2}{(2\pi)^{2}}\frac{\hbar\omega}{q^{2}}\right)\int d{\bf k}\ \frac{n_{{\bf k}}-n_{{\bf k}+{\bf q}}}{\hbar\omega+E_{{\bf k}}-E_{{\bf k}+{\bf q}}+i0^{+}},
\end{equation}

\begin{equation}
\sigma(\omega,q)=iG_{0}\hbar\omega\frac{1}{q^{2}}\frac{g_{s}g_{v}/2}{(2\pi)^{2}}\left(I_{1}(q,\omega)+I_{1}^{*}(q,-\omega)\right).
\end{equation}

Focusing now on $I_{1}(q,\omega)$, for a single isotropic massive
band $E_{{\bf k}}=\frac{\hbar^{2}\left|{\bf k}\right|^{2}}{2m}$,
and at null temperature and Fermi energy $\mu,$ $n_{{\bf k}}=\Theta(\mu-E_{{\bf k}})$,
and we have

\begin{equation}
I_{1}(q,\omega)=\int_{0}^{2\pi}d\theta\int_{0}^{\infty}kdk\ \frac{\Theta(\mu-E_{{\bf k}})}{\hbar\omega+E_{{\bf k}}-E_{{\bf k}+{\bf q}}+i0^{+}}.
\end{equation}
Defining $x=k/k_F$, and $y=q/k_F$, we have
\begin{equation}
I_{1}(q,\omega)=\frac{k_{F}^{2}}{\mu}~\int_{0}^{1}xdx\ \int_{0}^{2\pi}~\frac{d\theta}{\frac{\hbar(\omega+i0^{+})}{\mu}-y^{2}-2xy\cos\theta},
\end{equation}

The integral is easily done and yield

\begin{equation}
I_{1}(q,\omega)=2\pi\frac{k_{F}^{2}}{\mu}\frac{1}{2y}\left[\left(\frac{\frac{\hbar\omega}{\mu}-y^{2}}{2y}\right)-\sqrt{\left(\frac{\frac{\hbar\omega}{\mu}-y^{2}}{2y}\right)^{2}-1}\right]
\end{equation}
 for $\frac{\hbar\omega}{\mu}>(q/k_{F})^{2}$, whereas the case $\frac{\hbar\omega}{\mu}<(q/k_{F})^{2}$
is obtained by symmetry of the integral

\begin{equation}
I_{1}(q,\omega)=2\pi\frac{k_{F}^{2}}{\mu}\frac{1}{2y}\left[\left(\frac{\frac{\hbar\omega}{\mu}-y^{2}}{2y}\right)+\sqrt{\left(\frac{\frac{\hbar\omega}{\mu}-y^{2}}{2y}\right)^{2}-1}\right].
\end{equation}

\begin{figure}[ht]
\centerline{\includegraphics[width=8cm]{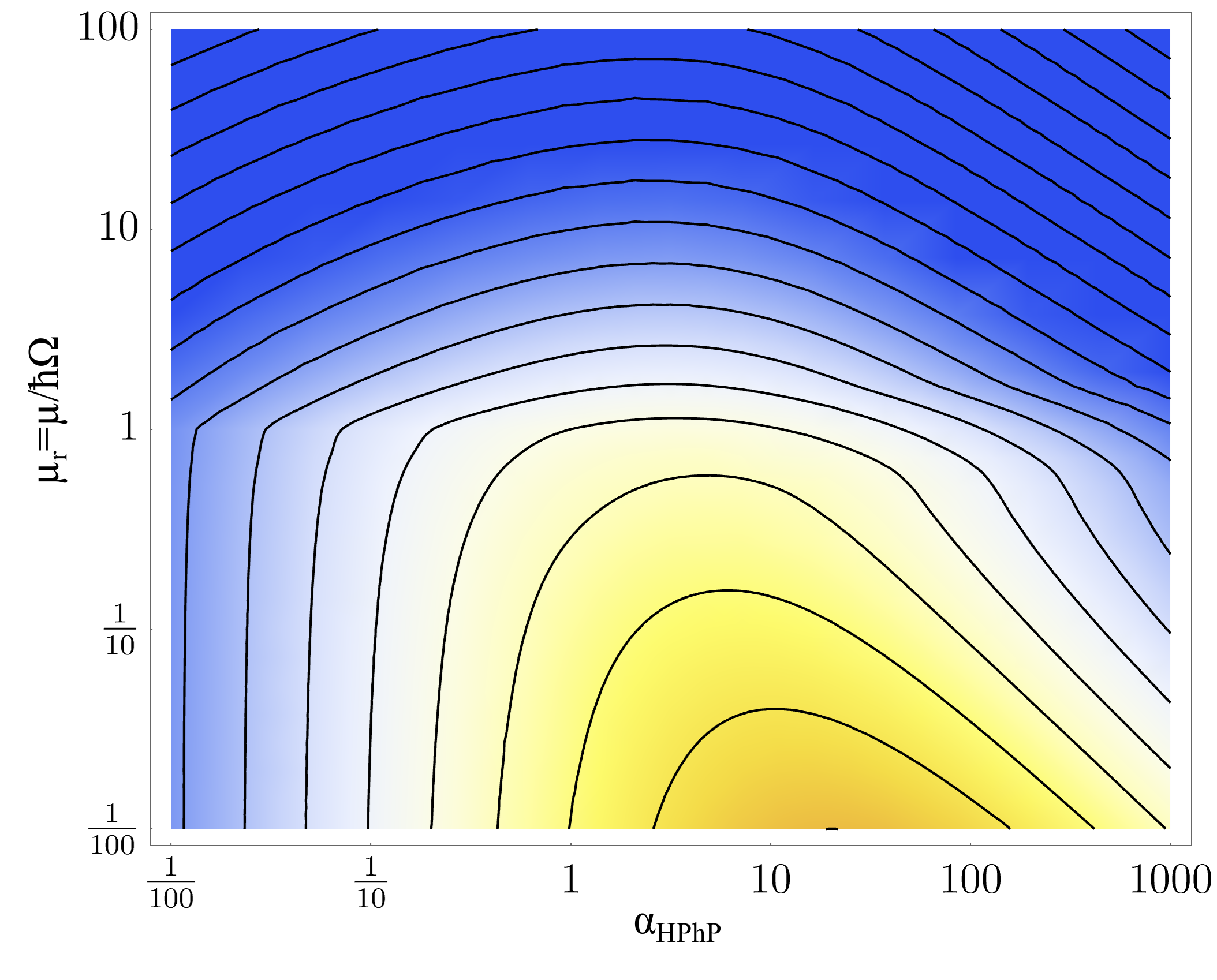}} \caption{SuperPlanckian material dependent factor at null temperature $A_{F}$ function of the HPhP fine structure constant $\alpha_{HPhP}$ and the reduced electronic chemical potential $\mu_r$. The prefactor is $A_F=0.023$ at $\alpha_{HPhP}=1$, $\mu_{r}=1$.}
\label{fig:AF}
\end{figure}

Figure \ref{fig:AF} represents the computed $A_{F}$ factor at null
temperature function of the channel/insulator properties. Corresponding
illustration of the nonlocal impedance matching factor is represented
on Figure \ref{fig:MatchingFactorNullTemp}.

The nonlocal conductivity in the degenerate case has been computed in detail for monolayer [\onlinecite{Principi2009prb}] and bilayer [\onlinecite{PrincipiThesis}] graphene and allows computing the radiated power using the above formalism (see [\onlinecite{Principi2017prl}] for the case of monolayer graphene).

\section{Acknowledgments}

The research leading to these results have received partial funding from the European Union ``Horizon 2020'' research and innovation program under grant agreement No. 785219 ``Graphene Core'', and from the ANR-14-CE08-018-05 ``GoBN''.


\begin{thebibliography}{10}

\bibitem{Pop2010nres} E. Pop, \emph{Nano Res.} \textbf{2010}, \emph{3}, 147. 

\bibitem{Ong20192DM} Z-Y. Ong, M-H. Bae, \emph{2D Mater.} \textbf{2019},\emph{6}, 032005. 

\bibitem{Yang2018nnano} W. Yang, S. Berthou, X. Lu, Q. Wilmart, A.
Denis, M. Rosticher, T. Taniguchi, K. Watanabe, G. F\`eve, J.M. Berroir,
G. Zhang, C. Voisin, E. Baudin, B. Pla\c{c}ais, \emph{Nat. Nanotechnol.} \textbf{2018}, \emph{13}, 47. 

\bibitem{Andersen2019science} T. I. Andersen, B. L. Dwyer, J. D. Sanchez-Yamagishi, J. F. Rodriguez-Nieva, K. Agarwal, K. Watanabe, T. Taniguchi, E. A. Demler, P. Kim, H. Park, M.D. Lukin, \emph{Science} \textbf{2019}, \emph{364}, 154. 

\bibitem{Tielrooij2018nnano} K.J. Tielrooij, N.C.H. Hesp, A. Principi,
M.B. Lundeberg, E.A.A. Pogna, L. Banszerus, Z. Mics, M. Massicotte,
P. Schmidt, D. Davydovskaya, D.G. Purdie, I. Goykhman, G. Soavi, A.
Lombardo, K. Watanabe, T. Taniguchi, M. Bonn, D. Turchinovich, C.
Stampfer, A. C. Ferrari, G. Cerullo, M. Polini, F.H.L. Koppens, \emph{Nat. Nanotechnol.} \textbf{2018}, \emph{13}, 41. 

\bibitem{Biehs2012prl} S-A. Biehs, M.Tschikin, P. Ben-Abdallah, \emph{Phys. Rev. Lett.} \textbf{2012}, \emph{109}, 104301. 

\bibitem{Biehs2014apl} S-A. Biehs, M.Tschikin, R. Messina, P. Ben-Abdallah, \emph{Appl. Phys. Lett.}\textbf{2014}, \emph{105},161902. 

\bibitem{Caldwell2014ncomm} J.D. Caldwell, A.V. Kretinin, Y. Chen,
V. Giannini, M.M. Fogler, Y. Francescato, C.T. Ellis, J.G. Tischler,
C.R. Woods, A.J. Giles, M. Hong, K. Watanabe, T. Taniguchi, S.A. Maier,
K.S. Novoselov, \emph{Nat. Commun.} \textbf{2014}, \emph{5}, 5221. 

\bibitem{Dai2015nnano}  S. Dai, Q. Ma, M. K. Liu, T. Andersen, Z. Fei, M. D. Goldflam, M. Wagner, K. Watanabe, T. Taniguchi, M. Thiemens, F. Keilmann, G. C. A. M. Janssen, S-E. Zhu, P. Jarillo-Herrero, M. M. Fogler, D. N. Basov, \emph{Nat. Nanotechnol.} \textbf{2015}, \emph{10}, 682. 

\bibitem{Kumar2015nl} A. Kumar, T. Low, K.H. Fung, P. Avouris, N.X.
Fang, \emph{Nano Lett.} \textbf{2015}, \emph{15}, 3172. 

\bibitem{Giles2016nl} A. J. Giles, S. Dai, O. J. Glembocki, A. V. Kretinin, Z. Sun, T. Chase, T. Ellis, J. G. Tischler, T. Taniguchi, K. Watanabe, M. M. Fogler, K. S. Novoselov, D. N. Basov, J. D. Caldwell, \emph{Nano Lett.} \textbf{2016}, \emph{16}, 3858. 

\bibitem{Low2017nmat}
T. Low, A. Chaves, J. D. Caldwell, A. Kumar, N. X. Fang, P. Avouris, T. F. Heinz, F. Guinea, L. Martin-Moreno, F. H.L. Koppens, \emph{Nat. Mater.} \textbf{2017}, \emph{16}, 182.  


\bibitem{Caldwell2019nrevmat}
\added{
J.D. Caldwell, I. Aharonovich, G. Cassabois, J.H. Edgar, B. Gil, D.N. Basov, \emph{Nat. Rev. Mater.}  \textbf{2019}, DOI:10.1038/s41578-019-0124-1 
}

\bibitem{Principi2017prl} A. Principi, M.B. Lundeberg, N.C.H. Hesp, K-J. Tielrooij, F.H.L. Koppens, M. Polini, \emph{Phys. Rev. Lett.}  \textbf{2017}, \emph{118}, 126804. 

 \bibitem{Santhanam2012prl}
P. Santhanam, D.J. Gray Jr, R.J. Ram,  \emph{Phys. Rev. Lett.}  \textbf{2012}, \emph{108}, 097403.  

\bibitem{Santhanam2013apl}
P. Santhanam, D. Huang, R.J. Ram, M.A. Remennyi, B.A. Matveev,  \emph{Appl. Phys. Lett.} \textbf{2013},  \emph{103}, 183513.   

\bibitem{Xue2015apl}
J. Xue, Y. Zhao, S.H. Oh, W.F. Herrington, J.S. Speck, S.P. DenBaars, S. Nakamura, R.J. Ram,  \emph{Appl. Phys. Lett.}  \textbf{2015}, \emph{107}, 121109.  

\bibitem{Liebendorfer2018xxx}
A. Liebendorfer,  \emph{AIP Scilight}  \textbf{2018}, DOI:10.1063/1.5037983   

\bibitem{Betz2012prl} A. C. Betz, F. Vialla, D. Brunel, C. Voisin, M. Picher, A. Cavanna, A. Madouri, G. F\`eve, J-M. Berroir, B. Pla\c{c}ais, E. Pallecchi, \emph{Phys. Rev. Lett.}  \textbf{2012}, \emph{109}, 056805. 

\bibitem{Betz2013nphys} A.C. Betz, S. H. Jhang, E. Pallecchi, R. Feirrera, G. F\`eve, J-M. Berroir, B. Pla\c{c}ais, \emph{Nat. Phys.}  \textbf{2013} \emph{9}, 109. 

\bibitem{Laitinen2014prb} A. Laitinen, M. Kumar, M. Oksanen, B. Pla\c{c}ais, P. Virtanen, P. Hakonen, \emph{Phys. Rev. B}  \textbf{2015}, \emph{91}, 121414 (R). 

\bibitem{Brunel2015jpcm} D. Brunel, S. Berthou, R. Parret, F. Vialla, P. Morfin, Q. Wilmart, G. F\`eve, J-M. Berroir, P. Roussignol, C. Voisin, \emph{J. Phys.: Condens. Matter}  \textbf{2015} \emph{27}, 164208. 

\bibitem{Crossno2016science} J. Crossno, J.K. Shi, K. Wang, X. Liu, A. Harzheim, A. Lucas, S. Sachdev, P. Kim, T. Taniguchi, K. Watanabe,
T.A. Ohki, K.C. Fong, \emph{Science}  \textbf{2016}, \emph{351}, 6277. 

\bibitem{Yang2018prl}  W. Yang, H. Graef, X. Lu, G. Zhang, T. Taniguchi, K. Watanabe, A. Bachtold, E.H.T. Teo, E. Baudin, E. Bocquillon, G. F\`eve, J-M. Berroir, D. Carpentier, M. O. Goerbig, B. Pla\c{c}ais, \emph{Phys. Rev. Lett.}  \textbf{2018}, \emph{121}, 136804. 

\bibitem{Nair2008science}
R.R. Nair, P. Blake, A.N. Grigorenko, K.S. Novoselov, T.J. Booth, T. Stauber, N.M.R. Peres, A.K. Geim, \emph{Science}  \textbf{2008} \emph{16}, 1308. 



\bibitem{Wurfel1982jpc}
P. Wurfel,   \emph{J. Phys. C: Solid State Phys.}  \textbf{1982},  \emph{15}, 3967.  

\bibitem{Rosencher2002cambridge}
E. Rosencher, B. Vinter,  \emph{Optoelectronics}, Cambridge University Press, \textbf{2002}.

\bibitem{Greffet2019prx} J.J. Greffet, P. Bouchon, G. Brucoli, \emph{Phys. Rev. X}  \textbf{2018}, \emph{8}, 021008. 


\bibitem{Mak2014apl}
K.F. Mak, C.H. Lui, T. Heinz,  \emph{Appl. Phys. Lett.}  \textbf{2014}, \emph{97}, 221904. 

\bibitem{Malic2017andp}
E. Malic, T. Winzer, F. Wendler, S. Brem, R. Jago, A. Knorr, M. Mittendorff, J. C. K\"{o}nig-Otto, T. Pl\"{o}tzing, D. Neumaier, H. Schneider, M. Helm, S. Winnerl,  \emph{Ann. Phys. (Berlin, Ger.)}  \textbf{2017}, \emph{529}, 1700038. 

\bibitem{Meric2008nnano}
\added{
I. Meric, M.Y. Han, A.F. Young, B. Ozyilmaz, P. Kim, K.L Shepard,  \emph{Nat. Nanotechnol.} \textbf{2008}, \emph{3}, 654. 
}

\bibitem{Bistritzer2011pnas} R. Bistritzer, A.H. MacDonald, \emph{PNAS}  \textbf{2011}, \emph{108}, 12233. 


\bibitem{Cao2018nature} Y. Cao, V. Fatemi, S. Fang, K. Watanabe, T. Taniguchi, E. Kaxiras, P. Jarillo-Herrero,  \emph{Nature}  \textbf{2018}, \emph{556}, 43. 


\bibitem{Yankowitz2019science} M. Yankowitz, S. Chen, H. Polshyn, Y. Zhang, K. Watanabe, T. Taniguchi, D. Graf, A.F. Young, C.R. Dean,  \emph{Science}  \textbf{2019}, \emph{363}, 1059. 



\bibitem{Banszerus2015sciadv} L. Banszerus, M. Schmitz, S. Engels, J. Dauber, M. Oellers, F. Haupt, K. Watanabe, T. Taniguchi, B. Beschoten, C. Stampfer, \emph{Sci. Adv.}  \textbf{2015},  \emph{1}, 1500222.  


\bibitem{Banszerus2016nl} L. Banszerus, M. Schmitz, S. Engels, M. Goldsche, K. Watanabe, T. Taniguchi, B. Beschoten, C. Stampfer, \emph{Nano Lett.}  \textbf{2016},  \emph{16}, 1387.  

\bibitem{Schue2016nscale} L. Schu\'e, B. Berini, A.C. Betz, B. Pla\c{c}ais, F. Ducastelle, J. Barjon, A. Loiseau, \emph{Nanoscale}  \textbf{2016}, \emph{8}, 6986. 


\bibitem{Schue2019prl}
L. Schu\'e, L. Sponza, A. Plaud, H. Bensalah, K. Watanabe, T. Taniguchi, F. Ducastelle, A. Loiseau, J. Barjon, \emph{Phys. Rev. Lett.}  \textbf{2019},  \emph{122}, 067401.  


\bibitem{Salihoglu2019jqsrt} H. Salihoglu, X. Xu, \emph{J. Quant. Spectrosc. Radiat. Transfer}  \textbf{2019}, \emph{222}, 115. 


\bibitem{Low2014small} C.G. Low, Q. Zhang, Y. Hao, R. S. Ruoff, \emph{Small}  \textbf{2014}, \emph{10}, 4213. 

\bibitem{Hu2018as} H. Hu, X. Guo, D. Hu, Z. Sun, X. Yang, Q. Dai, \emph{Adv. Sci.}  \textbf{2018}, \emph{5}, 1800175. 

\bibitem{Narimanov2015nphot} E.E. Narimanov, A.V. Kildishev, \emph{Nat. Photonics}  \textbf{2015}, \emph{9}, 214. 


\bibitem{Esslinger2014acsphot}     M. Esslinger, R. Vogelgesang, N. Talebi, W. Khunsin, P. Gehring, S. de Zuani, B. Gompf, K. Kern,  \emph{ACS Photonics}  \textbf{2014}, \emph{1}, 1285. 


\bibitem{Inhofer2018prapp} A. Inhofer, J. Duffy, M. Boukhicha, E. Bocquillon, J. Palomo, K. Watanabe, T. Taniguchi, I. Est\`eve, J-M. Berroir, G. F\`eve, B. Pla\c{c}ais, B.A. Assaf, \emph{Phys. Rev. Appl.}  \textbf{2018}, \emph{9}, 024022. 


\bibitem{Taur1997ieee} Y. Taur, D.A. Buchanan, W. Chen, D.J. Frank, K.E. Ismail, S-H. Lo, G.A. Sai-Halasz, R.G. Viswanathan, H-J.C. Wann, S.J. Wind, H-S. Wong, \emph{Proc. IEEE}  \textbf{1997}, \emph{85}, 486. 


\bibitem{Winta2019prb} C.J. Winta, M. Wolf, A. Paarmann, \emph{Phys. Rev. B}  \textbf{2019}, \emph{99}, 144308. 

\bibitem{Nyquist1928prb} H. Nyquist, \emph{Phys. Rev. B}  \textbf{1928}, \emph{32}, 110. 

\bibitem{Hooge1981rpp} F. N. Hooge, T. G. M. Kleinpenning, L. K. J. Vandamme, \emph{Rep. Prog. Phys.}  \textbf{1981}, \emph{44}, 479. 

\bibitem{Principi2009prb}
A. Principi, M. Polini, G. Vignale, \emph{Phys. Rev. B}  \textbf{2009}, \emph{80}, 075418. 

\bibitem{PrincipiThesis}
\added{
Alessandro Principi, \emph{PhD thesis}, Scuola Normale Superiore (Pisa, Italy) \textbf{2012}. 
}




\end{thebibliography}
\end{document}